# Fast transmission matrix measurement of a multimode fiber with common path reference

Raphael FLORENTIN, Vincent KERMENE, Agnès DESFARGES-BERTHELEMOT, Alain BARTHELEMY

XLIM Research Institute, University of Limoges, 123 av. A. Thomas, 87060 Limoges cedex France

Corresponding author: alain.barthelemy@xlim.fr

**Summary:** We report an improved scheme for the measurement of the transmission matrix of multimode waveguides in which the reference field co-propagates with the signal wave. The performance of the technique is demonstrated with the measurement of a 1.6 m long multimode optical fiber guiding 104 LP modes at 1064nm. The transmission matrix permitted efficient focusing of the light delivered at the fiber exit as well as shaping in the fiber's transmission channels.

Transmission matrix (TM) is a powerful tool for the control of light wave propagation through complex linear medium [1]. Knowledge of the TM makes possible the shaping of the beam delivered through a scattering medium [2] or through a multimode waveguide with coupling [3], it makes possible as well imaging through opaque medium [4]. On a more fundamental side, TM gives data on the nature of the light transport mechanism, it permits to derive the singular eigenmode input which leads to the highest power transmission [5]. Measured on a broad spectral range, TM allows computation of the time delay matrix and its Wigner-Smith eigenstates [6-7] opening the space-time control of the fiber output. A difficulty in the experimental determination of the TM comes from the measurement of the output optical field amplitude and phase associated to each element of the excitation base since a camera records the intensity pattern only. One standard way consists in the superposition of the beam leaving the complex medium with a plane wave provided by a reference arm [3,8,9]. A main issue with this configuration is the stability of the reference needed on the duration of the measurement. A reference-less option was also implemented based on binary amplitude modulated inputs combined with phase retrieval techniques [10]. Another reference less approach specific to multimode waveguides relies on multiple amplitude correlation with computer generated holograms of the waveguide eigenmodes [11]. In the present letter we show that TM measurement with a co-propagating reference can be efficient, fast and well suited to the characterization of a long piece of multimode optical fiber. Although TM of bulk scattering media (80μm thick layer of ZnO) has been measured in a co-propagating scheme [2,4] it is the first time that the technic is implemented on a long (> 1meter) multimode waveguide.

In most cases, the input field serving to probe the sample under test is shaped in phase only by means of a spatial light modulator (SLM). In order to provide a reference, the SLM surface is divided in two sections. One fraction only of the pixels serves for the generation of the probe field by spatial phase modulation. The second fraction is kept with a uniform phase. In previous works on scattering media the reference surface was chosen to cover 35% of the total SLM surface (disc shape) around a central square modulated area [2,4]. In the experiments reported below, it was reduced to 5% only of the spatial modulator which was a fast segmented deformable mirror (SDM) with 952 actuators (Boston Micromachines KiloDM). Furthermore the 50 elementary mirrors of the SDM which served to provide a reference were randomly distributed in the cross-section. For bulk samples, a reference output field is formed by scattering of the light coming from the reference surface. In the case of waveguide, the reference output field comes from the modes excited by the light reflected by the mirror reference elements, their propagation and coupling in the waveguide and their interference at the output. In the context of multimode optical fibers which are extremely sensitive to external perturbations, the common path reference ensures a crucial stability and accuracy in the measurement of the spatial phase delay. For each vector of the input basis (each phase chart), recovery of the phase of the output field was performed by the known technique of phase stepping interferometry. Even if the reference field is undetermined, provided it remains fixed on the duration of the recordings, setting the whole reference pixels on four different phase values successively gives four different intensity patterns from which one can compute the output field phase distribution. It was shown that the reference speckle does not impair the statistical properties of the TM and preserve the capability to focus or image using the measured TM [2,4].

As input basis we need an orthogonal set of input wavefronts. The number of SDM pixels is not compatible with the choice of a binary modulated Hadamard basis. Tilted plane wavefronts associated to a discrete sampling of the wave-vector space is a possible choice. Our computations show however that the number of excited MMF eigenmodes strongly evolves when scanning the input basis. As an alternative quasi orthogonal basis, it is possible to use a set of 900 random phase distributions (for the 900 SDM elements used for wavefront shaping). The beam from a CW fiber laser operating at 1064nm was first expanded and collimated to cover the 10mm diameter of the SDM (see Fig. 1). The modulated beam was further demagnified and imaged onto the input facet of the MMF. The MMF was 1.6 m long step index fiber, with a core diameter of 90µm and it carried 104 modes per polarization. The fiber was loosely wounded on the optical table. The output figure of the MMF was imaged with magnification onto a 16bits CCD camera (FLIR Grasshopper 3). The smallest speckle grain expected on the fiber output facet is of the order of λ/2NA~5µm so that the fiber core diameter represents about 20 times this value. The 128x128 pixels images taken from the CCD camera were therefore fully sufficient for a good sampling and phase recovery of the output field. The 4x900 recorded images were then processed to get the transmission matrix between the input phase pixels (M=900) and the pixels in the output image (N=16384=128x128) of the MMF output intensity. One TM measurement takes less than 4 minutes, a limit set by the non-optimized transmission and storing speed in the computer memory (15 frames/s.)). In order to evaluate the quality of the measured TM we computed the focusing operator which indicates the capability of the system to form a sharp spot (a focus) on any pixel of the output cross section. Given the fact that we can shape the input phase profile only and not its amplitude the focusing operator is given by:

$$O_{foc} = TM . TM^{\dagger}_{norm} \quad \text{[Equ.1]}$$

with the element of $TM_{norm}$ being normalized to their module $a_{i,j}/|a_{i,j}|$ and where the $\dagger$ symbol denotes the conjugate transpose. Figure 2 shows the focusing operator of our system as computed from the measured TM after down sampling to 1024 pixels to make the display more readable. The strong peak (yellow line) on the diagonal means that the focusing should be efficient on each pixel of the fiber cross section. Because the pixel in the image is of the order of 3.3µm, i.e. smaller than the smallest speckle grain (~5µm), the focus spot which can be shaped covers a surface a little bit larger than a single pixel. That is the reason why one can see in Fig.2 a parallel line on each side of the diagonal. Their intensity is weak but sufficient to make them visible on top of the background.

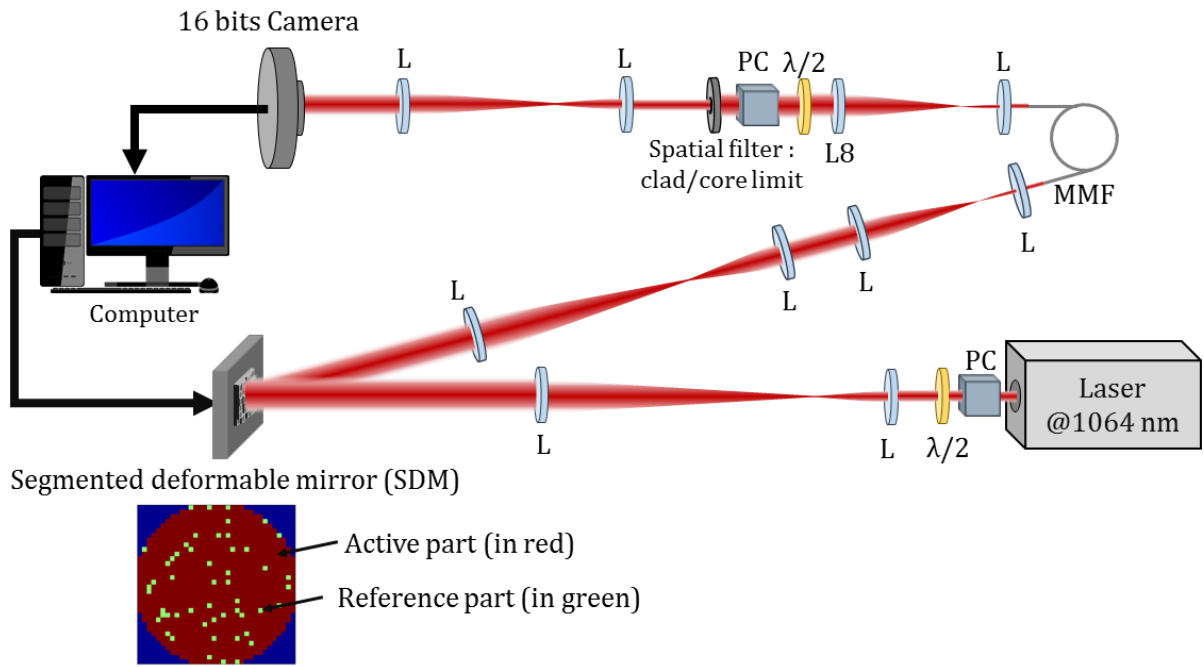

**Figure 1** : Experimental set up (L: lens, PC: polarizing cube, MMF: multimode fiber). The inset shows an example of distribution of the pixels on the SDM for respectively the reference (green) and shaped input wavefront (red).

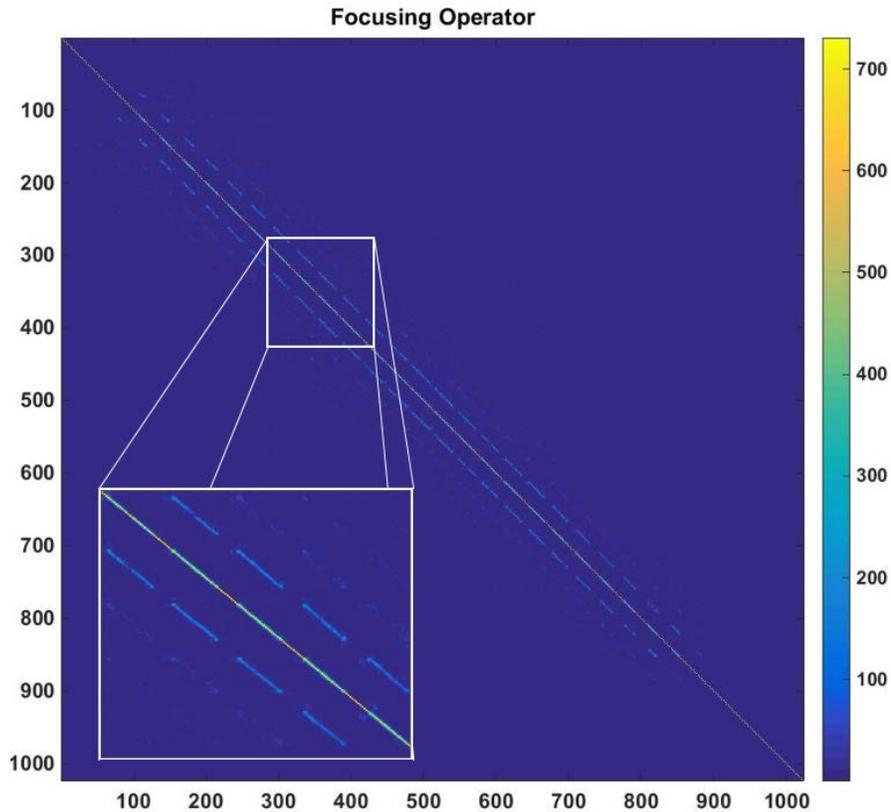

*Figure 2*: Focusing operator on 1024 positions, computed from the measured transmission matrix of the MMF with a zoom on the diagonal in inset

We have checked the use of the TM for focusing through the 1.6 meter long MMF. To get a desired output field $E_{out}$, the input phase profile must be shaped according to:

$$E_{in} = TM^{\dagger}.E_{out} / |TM^{\dagger}.E_{out}| \quad \text{[Equ.2]}$$

Three typical output patterns are presented on Fig. 3. Fig. 3-a was obtained with a random input wavefront and serves for reference Fig.3 b and c show focusing of the output beam in two different areas when the SDM shaped the input wavefront according to Equ.2.

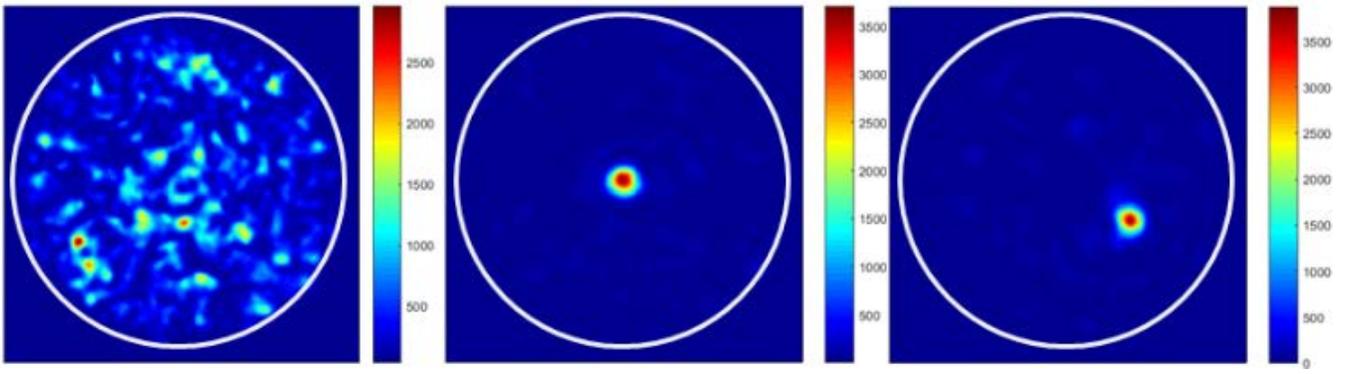

*Figure 3: Output intensity pattern obtained for a random input wavefront (a), for a wavefront shaping based on the TM to get a focus at the fiber center (b) and then to get a focus in an off-centered area (c).*

Beam focusing through a complex medium is usually assessed by the enhancement factor $\eta$ given by the ratio between the peak intensity on the focus spot $I_{foc}$ and the average value of the background $\langle I \rangle$: $\eta = \frac{I_{foc}}{\langle I \rangle}$. We measured here an intensity enhancement value of 80.7 for a centered focus and 75.1 for the off-centered one. These values are extremely close to the value expected from theory [12] $\eta = \frac{\pi}{4}.(N-1)+1 = 82$ (N denoting the number of freedom degrees, here given by the number of modes).

Because the SDM was a fast shaping device, once the TM has been recorded, it is possible to perform a raster scanning of the output focus at a 30 kHz speed. In a second step we computed the singular value decomposition (SVD) of the TM. In order to demonstrate a more complex spatial control of the MMF output, we chose to shape the output field according to some singular vectors of the SVD, also known as transmission channels. Taking the output singular vectors corresponding to some of the ten stronger singular values as successive targets, we displayed on the SDM the wavefront of the associated input singular vectors. We provide on Fig. 4 the five theoretical patterns of the singular vectors and the five patterns experimentally recorded at the fiber output after wavefront shaping derived from the TM SVD.

The observed output figures look very close to the expected ones although the shaping of the excitation was achieved on the phase front only, keeping an overall Gaussian intensity distribution.

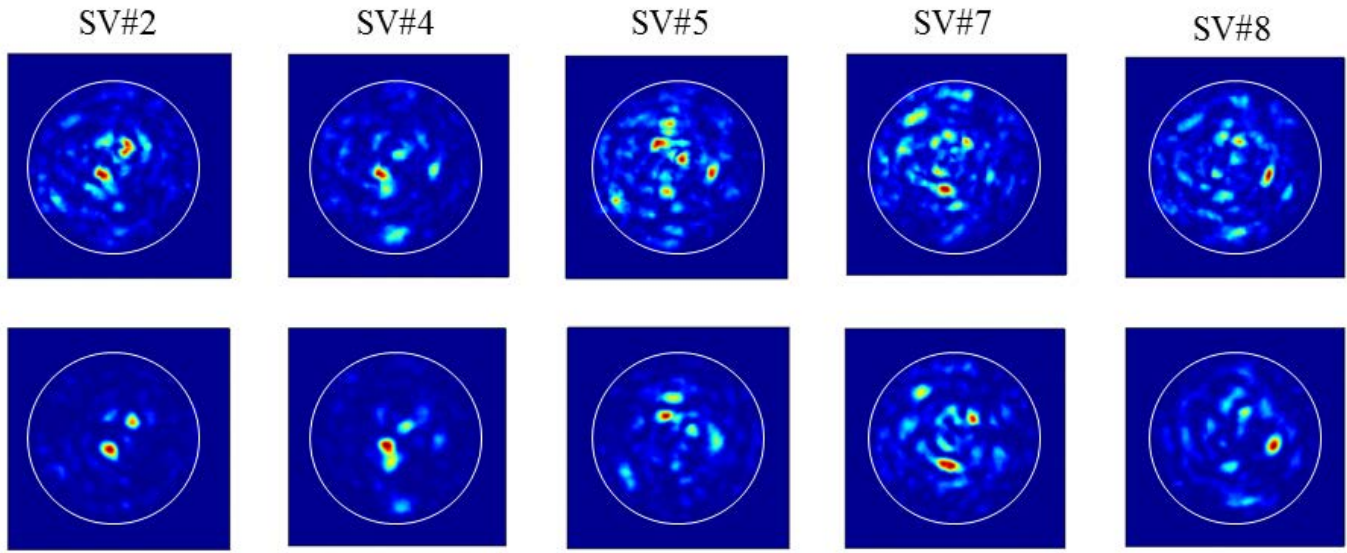

*Figure 4: Comparison between the theoretical output singular vectors computed from the SVD of the measured TM and the experimental patterns observed after shaping of the MMF input according to the input singular vectors. A few selected singular vectors are illustrated here chosen among the ones with the ten highest singular values.*

In conclusion, we have adapted to multimode waveguides the TM measurement scheme using a co propagating reference. We have demonstrated that 5% of the input wavefront devoted to the reference wave is enough to preserve the dynamic required for a complex value TM measurement. Based on the TM, focusing of the laser light transmitted through a 1.6 m long MMF was achieved with performances very close (91%-98%) to the theoretical limit. The common path benefits are a simplified set-up and an improved stability for the coherent interference with the reference. The techniques has been used to characterize a MMF amplifier at various gain levels where thermal effects due to the powerful pump laser were significant. Amplified beam control by use of the active fiber TM has been demonstrated and will be reported in a future paper. With an improved transmission and recordings of the output images during the measurement (500 frames/s. have ben already reported [7]) it should be possible to get the fiber TM in ~7 seconds and even faster with the acceleration provided by GPU [13].

**Acknowledgment**. We acknowledge funding from the Agence Nationale de la Recherche, POMAD project # ANR-14-CE26-0035-01